# Binary craters on Ceres and Vesta and implications for binary asteroids

C. Herrera[1,2], B. Carry[1], A. Lagain[3,4,5], and D. E. Vavilov[6]

[1] Université Côte d'Azur, Observatoire de la Côte d'Azur, CNRS, Laboratoire Lagrange, Nice, France
e-mail: `carianna.herrera@dlr.de` (C. Herrera), `benoit.carry@oca.eu` (B. Carry)
[2] Master track in Astrophysics, Université Côte d'Azur & Observatoire de la Côte d'Azur, Parc Valrose, 06100 Nice, France
[3] Space Science and Technology Centre, School of Earth and Planetary Sciences, Curtin University, Perth, WA, Australia.
e-mail: `lagain@cerege.fr`
[4] Aix Marseille Univ., CNRS, IRD, INRA, CEREGE, Aix en Provence, France.
[5] Aix-Marseille Univ., Institut ORIGINES, Marseille, France.
[6] Institute of Applied Astronomy, Russian Academy of Sciences, Kutuzova emb. 10, St. Petersburg, Russia.
e-mail: `vavilov@iaaras.ru`

**ABSTRACT**

*Context.* Airless planetary objects have their surfaces covered by craters, and these can be used to study the characteristics of asteroid populations. Planetary surfaces present binary craters that are associated with the synchronous impact of binary asteroids.
*Aims.* We identify binary craters on asteroids (1) Ceres and (4) Vesta, and aim to characterize the properties (size ratio and orbital plane) of the binary asteroids that might have formed them.
*Methods.* We used global crater databases developed in previous studies and mosaics of images from the NASA DAWN mission high-altitude and low-altitude mapping orbits. We established selection criteria to identify craters that were most likely a product of the impact of a binary asteroid. We performed numerical simulations to predict the orientation of the binary craters assuming the population of impactors has mutual orbits coplanar with heliocentric orbits, as the current census of binary asteroids suggests. We compared our simulations with our survey of binary craters on Ceres and Vesta through a Kolmogorov-Smirnov test.
*Results.* We find geomorphological evidence of 39 and 18 synchronous impacts on the surfaces of Ceres and Vesta, respectively. The associated binary asteroids are widely separated and similar in diameter. The distributions of the orientation of these binary craters on both bodies are statistically different from numerical impact simulations that assume binary asteroids with coplanar mutual and heliocentric orbits.
*Conclusions.* Although the identification of binary craters on both bodies and the sample size are limited, these findings are consistent with a population of well-separated and similarly sized binary asteroids with nonzero obliquity that remains to be observed, in agreement with the population of binary craters identified on Mars.

## 1. Introduction

Asteroids with natural satellites, also referred to as binary asteroids, represent a significant fraction of the population of impactors in the inner Solar System. It is estimated that $15 \pm 4\%$ of small asteroids (diameter of a few hundred meters to several kilometers) possess a satellite (Margot et al., 2002; Pravec et al., 2006), while only a couple of percent of the largest asteroids have moons (Margot et al., 2015).

The detection and characterization of satellites of small asteroids mainly rely on two observing techniques: radar echoes for near-Earth asteroids (NEAs) (Benner et al., 2015) and optical light curves for both NEAs and main belt asteroids (MBAs) (Pravec et al., 2006). While both techniques are efficient and precise, they suffer from biases, including a limited range from the Earth for radar observation (mostly limiting the sample to NEAs) and a strong preference for compact and low-obliquity systems for light curves (linked to the detection from mutual eclipses and occultations).

A substantial portion of binary systems with a diameter of below 10 km share highly similar properties: rapidly rotating primaries, an obliquity of close to 0°or 180°, a secondary-to-primary diameter ratio of $d_s/d_p \approx 0.3$, and a system-semi-major-axis-to-primary-diameter ratio of $a/d_p \approx 2$ (see Margot et al., 2015; Minker & Carry, 2023). These characteristics are in line with the recent discovery of a second asteroid orbiting Dinkinesh during the NASA Lucy spacecraft flyby on November 1, 2023, and the Didymos-Dimorphos system, which is target of the NASA DART and ESA Hera missions (Levison et al., 2017; Rivkin et al., 2021; Michel et al., 2022).

Recently, Vavilov et al. (2022) conducted a survey for binary craters on the surface of Mars thought to be formed by the impact of binary asteroids. These authors found evidence for a population of binary asteroids with different properties: components are widely separated, similar in size, and not limited to low obliquity. Although the results of this survey are strongly biased due to limitations in identifying such structures via geomorphological evidence, the characteristics of Martian binary craters suggest the existence of a population of binary asteroids exhibiting still unknown size and orbital characteristics.

The aim of the present study is to test these recent findings by exploring the properties of binary craters on different planetary surfaces. We focus on the two asteroids Ceres and Vesta, targets of the NASA DAWN mission (Russell et al., 2012a, 2016). Preliminary efforts to identify binary craters on these two asteroids have been made (e.g., Wren & Fevig, 2018, 2023), although these studies covered a restricted area (∼3% and 15% of the surface of Vesta and Ceres, respectively) and were limited to craters of >3 km in diameter. In the present study, we established criteria of identification, performed a global survey of both bodies, and analyzed the binary crater population in terms of their size ratio and orientation.

The article is organized as follows. In Sect. 2 we describe the imagery dataset used for both Ceres and Vesta. In Sect. 3 we describe the method and criteria used to identify binary craters. In Sect. 4 we present both catalogs of binary craters with their properties. In Sect. 5 we present the simulation of known binary asteroid impacts used to compare with the binary craters we identified on the surface of Ceres and Vesta. In Sect. 6 we finally compare our results with previous findings on Mars and on some restricted areas of Ceres and Vesta, and discuss the implications for the population of binary asteroids in the main belt.

## 2. Data sources

We used the High- and Low-Altitude Mapping Orbit (HAMO and LAMO) images from the DAWN mission to find geomorphological evidence of synchronous impact craters on the surfaces of Ceres and Vesta (Russell et al., 2015). Both imagery datasets were assembled into global mosaics and are available on the Planetary Data System (PDS). For Ceres, the HAMO mosaic has a resolution of 140 m/px and the LAMO mosaic has a resolution of 35 m/px, while for Vesta, the resolutions of the mosaics are 60 m/px and 20 m/px, respectively.

We used the global crater databases of Zeilnhofer & Barlow (2021b) for Ceres and the one developed by Liu et al. (2018) for Vesta as a starting point. These databases contain a total of 44,594 craters with diameters of greater than 1 km for Ceres and 11,604 craters with diameters of greater than 700 m for Vesta. All craters in both catalogs were considered in this study and surveyed with the Arcmap ESRI software. The imagery dataset was also reprojected using stereographic projections for high latitudes in order to avoid geometrical distortions of impact craters.

For both bodies, the HAMO and LAMO mosaics both have near-global coverage. Although we mostly relied on LAMO mosaics, given that their resolutions are about three to four times higher, there were some regions with images presenting artifacts and of poor quality in general. For Vesta, this results in a total of 1,301 craters (∼11% out of the total of craters in the Vesta catalog given by Liu et al., 2018) that were only surveyed using the HAMO mosaic, despite coarser image resolution. This is because the DAWN mission LAMO phase occurred during northern winter, which led to the north pole being in the dark. According to Roatsch et al. (2015), only 84% of the surface was illuminated, and the portion of the surface with appropriate illumination for crater identification is reduced to 66.8%. Another limitation of the LAMO mosaic of Vesta comes from the lack of data in the form of several diagonal strips. There were <0.5% craters located within these gaps, all of which were identifiable using the lower-resolution mosaic.

For Ceres, one of the defects in its corresponding LAMO images consists of imagery artifacts within an area of about 996 km$^2$ between 55-58°N and 105-110°E. Despite the resolution loss in using the HAMO mosaic covering this region, there were no evident pairs of craters that could have been misidentified.

Due to these limitations, our survey on both bodies results in a spatial incompleteness of the binary crater population. However, as discussed in the following section and further detailed in Vavilov et al. (2022), the main limiting factor in the identification of such craters is the conservation of morphological evidence attesting to a synchronous impact.

## 3. Criteria for identification of binary craters

The impact of binary asteroids on planetary surfaces may lead to a wide diversity of craters morphology, mostly depending on the size of the two impactors, their distance and mutual orientation with respect to the surface at the moment of the impact, and the physical characteristics of the target and the impactors (e.g., composition, density; see Miljković et al., 2013). Also, it is important to note that without precise radiometric dating of samples representative of the formation time of two impact craters, we can only assess the likely synchronicity of two impact events on the basis of morphological observations (see e.g., the nonsynchronicity of the Clearwater Lake impact structures evidenced by Ar-Ar radiometric dating by Schmieder et al., 2015). Here, we list the criteria we based our survey on to distinguish single from likely binary craters.

As opposed to the craters on Mars, most craters on Vesta and Ceres do not exhibit a continuous and thick ejecta blanket (Liu et al., 2018; Robbins & Hynek, 2012; Zeilnhofer & Barlow, 2021b; Lagain et al., 2021b), because their surface composition is significantly less hydrated and the atmospheric pressure is inexistent (Costard, 1989; Prettyman et al., 2012, 2017; Denevi et al., 2012; Bland et al., 2016). This limits the identification of synchronous impact craters that are far from one another, where the ratio between separation and main crater diameter is typically higher than one (Miljković et al., 2013). Therefore, the main morphological characteristic used in this study is the presence of a straight and continuous contact between the two craters (hereafter referred to as a septum), without any evidence of a stratigraphic relationship (Vavilov et al., 2022). As a further assessment of the synchronicity of two impact events, we also accounted for the difference in the preservation state of the impact structures for each potential pair. Specifically, we focused on any difference between the depth/diameter ratio of each pair, which indicates the relative age of an impact. When not reported in crater databases, crater depth is estimated visually based on the extent of the shade within the crater. Any significant difference in the age of the impacts constituting a pair of craters would result in a significant difference in the depth/diameter ratio and thus their preservation state (Stopar et al., 2017).

We also analyzed the surrounding terrains at a regional scale in order to exclude possible secondary craters from another (larger) impact. As these craters are formed at the same time as the primary crater from the fallback of distal ejecta on the surrounding surface, and locally increase the crater density (Robbins & Hynek, 2014; Lagain et al., 2021b,a), these structures can potentially share mutual characteristics with binary craters (similar preservation state and presence of a septum characterizing the contact between two craters). Thus, areas in the immediate vicinity of larger craters showing abundant secondaries are excluded from the survey.

A specific complexity arises in the case of Ceres, where many craters do present polygonal rims (Zeilnhofer & Barlow, 2021a). Polygonal-shaped craters on Ceres could be confused with a pair joined by a septum. We thus scrutinize the close vicinity of primary craters for the presence of other polygonal craters and avoid adding them to our catalog.

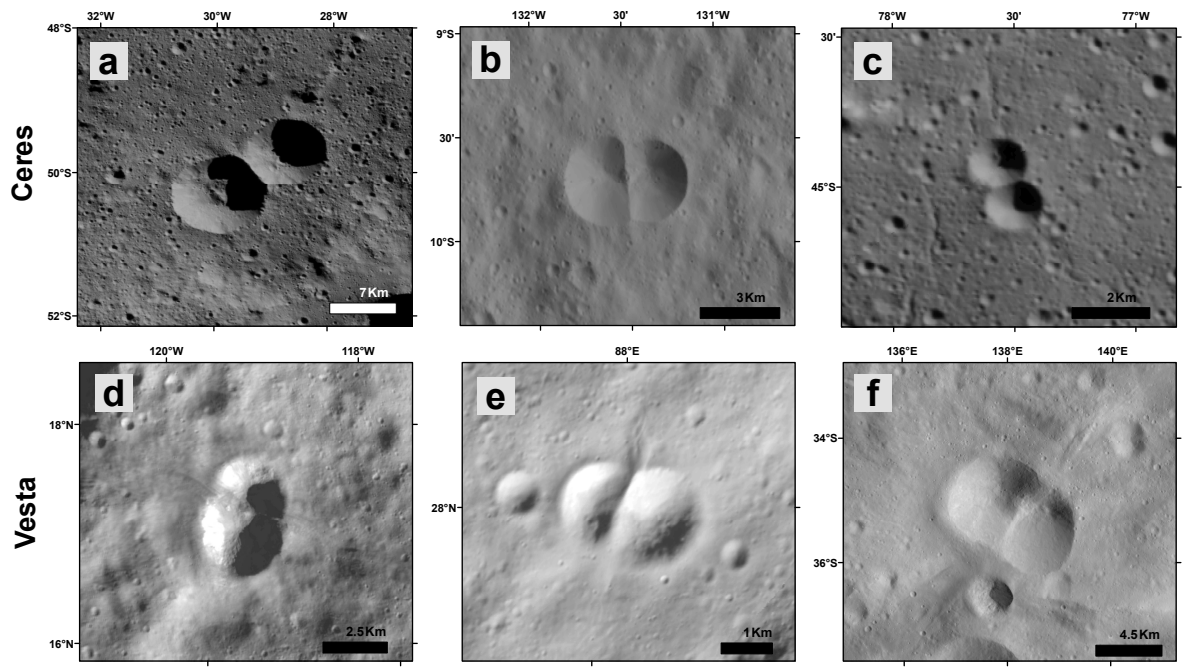

**Fig. 1.** Examples of binary craters identified on the surface of Ceres (top) and Vesta (bottom). The contact between all rim pairs is characterized by a continuous septum without any visible stratigraphic relationship, as well as a similar preservation state, thus indicating a very likely synchronous formation. We note that in the case of the pairs on Vesta presented here, an excess of ejecta material is visible in the direction of the septum, which is expected in the case of a binary asteroid impact (Miljković et al., 2013). Background imagery: LAMO mosaic.

The databases of binary craters we present in this study respect all morphological criteria listed above (septum, similar preservation state, not located in secondary crater fields, and not exhibiting polygonal rims). As the criteria can exhibit more or less marked morphological expressions, we assign a level of confidence of each binary crater: very likely and likely (respectively, score 2 and score 1 in Appendix A). Nevertheless, all values of binary craters are considered in the statistical analyses, and this distinction is simply presented in a comparative manner.

The location of some crater centroids on Ceres presents an offset of between hundreds of meters and a few kilometers between the database provided by Zeilnhofer & Barlow (2021b) and the imagery dataset we used. Therefore, we remeasured the diameter and location of all craters identified as binary on the LAMO images using the ESRI ArcMap CraterTools package (Kneissl et al., 2011). We performed these measurements by placing three points along the crater rim, following the method also used by Liu et al. (2018) for Vesta. We then measured the distance and the orientation angle for all pairs of craters using the updated centroid location.

We also computed the diameter ratio between the smaller and larger crater, and the separation ratio as the distance between craters normalized by the diameter of the larger crater. Finally, we assigned a morphological class to each binary according to the classification proposed by Miljković et al. (2013) based on their diameter ratio and separation. This classification was also used in a previous study that produced a survey of Martian binary craters (Vavilov et al., 2022). The classes are mentioned here and we refer the reader to Vavilov et al. (2022) for further details and examples:

- Doublet: The two craters do not show any contact between their rims, and the ejecta blanket morphology enables the recognition of a potential doublet. An excess of ejecta material perpendicular to the axis defined by the two crater centroids is visible between the two ejecta blankets.
- Peanut: the two craters are of approximately the same size, and their rims are in contact (septum) but are sufficiently separated to identify two distinct impact structures.
- Overlapping: a small impact crater is close to the large binary component and does not exhibit a circular morphology, suggesting an overlap of material ejected from the main crater onto the minor crater following the impact event.
- Tear: a minor impact structure is very close to the larger crater of the binary and there is no clear evidence of a stratigraphic relationship.
- Elliptical: the impact crater is elliptical and the main direction of its ejecta blanket is perpendicular to the major axis of the crater cavity.
- Circular: Some impact conditions —typically with small separation of both impactors and/or small diameter ratio— do not theoretically allow the formation of a binary crater (Miljković et al., 2013). However, we mark circular structures exhibiting a septum located on the central peak or a partially circular structure on the rim as potential binary craters.

We note that a low confidence level and a low number of binary craters falling in the three last categories are anticipated due to the fact that the morphological evidence characterizing these classes is more subtle. Some examples of the morphologies of the binary craters on Ceres and Vesta are shown in Fig. 1 and examples of morphologies not recognised as binary craters are shown in Fig. 2.

## 4. Description of binary craters on Ceres and Vesta

We identified 39 binary craters on Ceres and 18 on Vesta (Fig. 3). Among them, 17 and 5 binary craters were considered to be very likely formed by a binary asteroid impact on Ceres and Vesta, respectively (see tables A.1 and A.2). The complete sample is presented in Appendix A, which lists their longitude, latitude, diameter, separation, orientation, confidence degree index, and morphological class according to Miljković et al. (2013). Furthermore, we report the size of the impactor corresponding to each crater, assuming C-type and S-type asteroids.

As in the case of Mars (Vavilov et al., 2022), the spatial distribution of binary craters on both bodies does not present any pattern beyond that of the whole crater population on the surface (e.g., there is an heterogenous distribution of craters due to resurfacing: on Vesta this is typically due to the impacts that created the Venneneia and Rheasilvia bassins; Marchi et al., 2012).

Figure 4 shows the distribution of the diameter ratio of the binary craters as a function of the separation between the two components. The distribution of the diameter ratio and separation for known binary asteroids is also presented in a separate panel. The difference between binary asteroids and binary craters in terms of these parameter distributions is striking. However, close-in binary ($a/d_p \lesssim 0.3$) and/or small satellites ($d_s/d_p \lesssim 0.4$) are not expected to be recognized as binary craters (Miljković et al., 2013) and our survey is thus biased towards the recognition of impacts formed by widely separated binary asteroids where the two objects are of similar size.

The surface of Vesta is less hydrated than that of Ceres (Prettyman et al., 2012, 2017); compared to Mars, the recognition of widely separated craters is therefore much less straightforward (Vavilov et al., 2022). This explains the limited range of separation detected on Vesta —as shown in Fig. 4— and the overall absence of doublet craters identified in the present study. The separation-to-main-crater-diameter ratio on Mars goes up to 1.2 (Vavilov et al., 2022), while in this study only one binary crater had a ratio barely higher than 1 (exactly 1.03).

We present the orientation of the binary craters in Fig. 5. The orientation is defined as the angle between the equator and the line connecting the center of the two craters. We find that the orientation angles for both bodies cover mostly all angles and do not show a trend or preferential orientation, similar to the distribution of binary craters on Mars (Vavilov et al., 2022).

Additionally, we estimate the diameter of the impactor that might have formed each crater identified in our surveys for Ceres and Vesta. The impactor size $L$ is derived from the scaling laws given in Collins et al. (2005):

$$D_t = 1.161 \left(\frac{\rho_i}{\rho_t}\right)^{\frac{1}{3}} L^{0.78} v_i^{0.44} g^{-0.22} sin^{\frac{1}{3}}(\theta), \quad (1)$$

where $D_t$ is the diameter of the transient crater, $\rho_i$ and $\rho_t$ are respectively the density of the impactor (either C- or S-type impactors) and the target, $v_i$ is the impact velocity at the surface, $g$ the gravitational acceleration of the body, and $\theta$ the impact angle. The numerical values of these parameters are summarized in Table 1.

The computed value of the diameter of the transient crater $D_t$ is then used to estimate the final crater diameter ($D_f$):

$$D_f = \begin{cases} 1.17 \dfrac{D_t^{1.13}}{D_c^{0.13}}, & \text{if } D_t < D_c, \quad (2) \\ 1.25\, D_t, & \text{if } D_t > D_c, \quad (3) \end{cases}$$

where $D_c$ is the transition diameter between simple and complex craters (see Table 1).

The diameter of binary craters, $D_m$, as measured on Ceres from the imagery data and compared to the catalogs varies between ~1.1 km and ~33.8 km, while on Vesta this diameter ranges from ~0.9 km to ~13.0 km. The impactor diameter is then obtained by inverting the diameter of each crater of the survey. Our calculations suggest that identified binary impact craters on Ceres were produced by binary asteroids with diameters of between ~50 m and ~5 km (see Table A.3), while craters on Vesta were produced by binary asteroids ranging between ~50 m and ~1.2 km (see Table A.4). In comparison, Vavilov et al. (2022) estimated that binary impact craters on Mars were produced by binary asteroids ranging between ~70 m and ~3.8 km. Although there is a slightly wider range of binary impactor diameters for Ceres than for both Mars and Vesta, the maximum impactor size for Ceres is smaller than that for Mars, and these three bodies sample a population of binary asteroids of similar size.

## 5. Simulations of impacts

We simulated the impact of binary asteroids to compare the geometry of the resulting impacts with the observed cratering record. We only briefly summarize the approach here and refer to Vavilov et al. (2022) for a detailed description.

We considered all asteroids with a minimal orbital intersection distance (MOID, Marsden, 1993) with the orbits of Ceres and Vesta of smaller than 0.05 AU. Asteroid orbits are taken from JPL[1]. We assume that all these asteroids are binaries and that the collision takes place at the MOID point, which is the location in space where the orbits of the binary and Ceres or Vesta are the closest to each other.

Following the current census of binary asteroids (Margot et al., 2015), we assume here that the mutual orbit of each binary asteroid system is circular and coplanar with its heliocentric orbit (i.e., their obliquity is null). We consider 360 possible mutual positions of the two components of the binary and project them all onto the target plane (Kizner, 1961). Then, we compute the angle between the equator of the target body (Ceres or Vesta) and the line connecting the two centers of the projectiles on this plane. We use the spin-vector coordinates of Ceres and Vesta listed in Table 1.

This approach is simplified and does not take into account the gravitational focusing from Ceres and Vesta, their shape, or the long-term evolution of their spin axis. Both Melosh & Stansberry (1991) and Vavilov et al. (2022) showed that neither the separation nor the orientation of the binary system is affected during an impacting trajectory to the Earth or Mars by gravitational focusing. We performed a similar full dynamical simulation on Ceres and Vesta. The difference between the simplified and full-dynamics solutions is indeed negligible: it does not exceed 1°, which is smaller than the accuracy of our observations.

The obliquity of both Ceres and Vesta has evolved over time, and excursions in the ranges of 2–20° and 21–45° have been predicted for both asteroids (Vaillant et al., 2019). For Mars, Vavilov et al. (2022) accounted for the changing obliquity by summing different realizations of the simulations with different obliquities, weighted by the time of residence in each spin state (taken from Laskar et al., 2004). As the distribution of time of residence is not available for Ceres or Vesta, we cannot simulate it. However, the effect of a varying obliquity is qualitatively simple: it diminishes the contrast of the distribution of orientations (Fig. 5).

---

[1] http://ssd.jpl.nasa.gov/sbdb_query.cgi

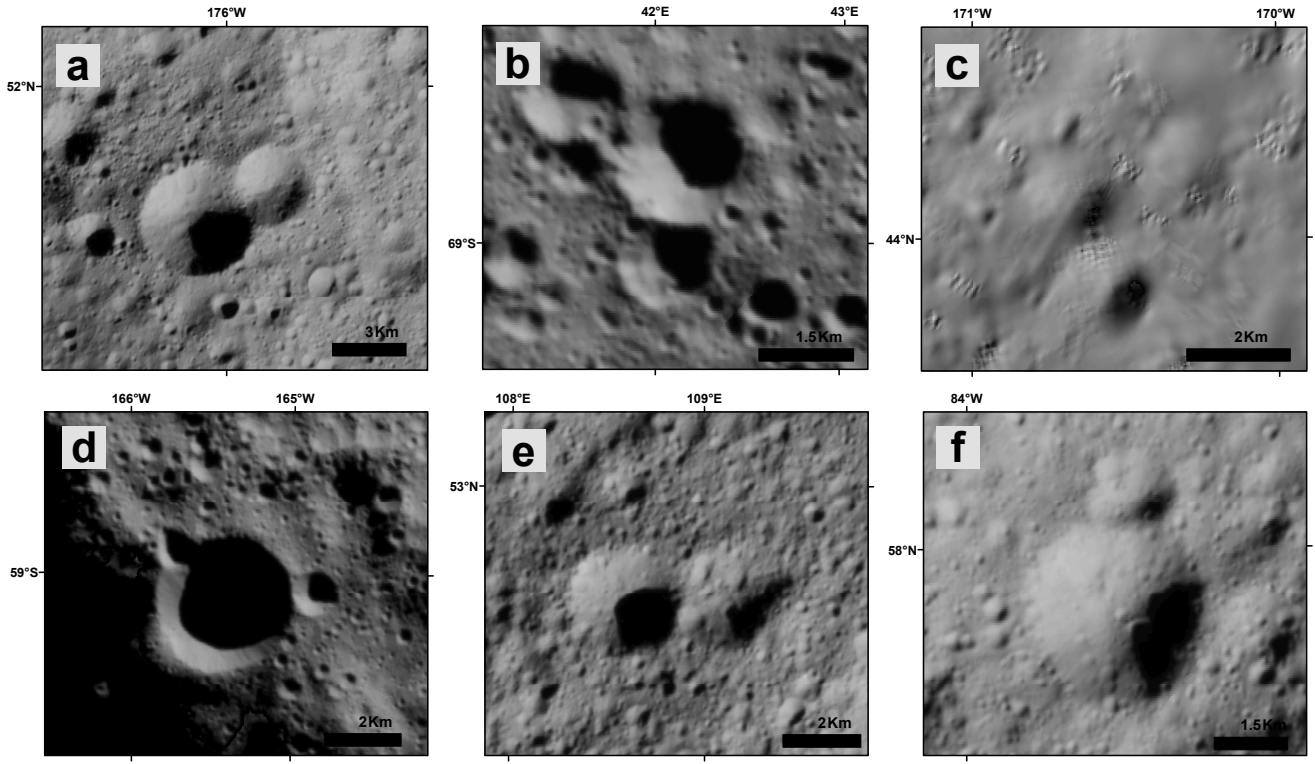

**Fig. 2.** Example of craters on Ceres not recognized as binary. Clear stratigraphic relationships can be established in the case of craters presented in panels (a), (b), (d), and (f); typically the rim in contact between the craters is not a septum but is characterized by a circular shape. In the case of panel (c), the imaging data present artifacts that preclude any conclusion regarding the synchronicity of the impacts. Panel (e) shows two craters with a septum but their degradation state and depth/diameter ratio are significantly different, indicating a nonsynchronous impact.

**Table 1.** Physical parameters of Ceres, Vesta, and the impactors used in this study.

| Body | Parameter | Symbol | Units | Value | Reference |
|---|---|---|---|---|---|
| Ceres | Right Ascension of the spin axis | $\alpha_0$ | deg. | $291.42 \pm 0.01$ | Russell et al. (2016) |
| | Declination of the spin axis | $\delta_0$ | deg. | $66.76 \pm 0.02$ | Russell et al. (2016) |
| | Mean radius | $R$ | km | $469.7 \pm 0.2$ | Russell et al. (2016) |
| | Mean density | $\rho$ | kg.m$^{-3}$ | $2161.6 \pm 2.5$ | Park et al. (2019) |
| | Surface gravity | $g$ | m.s$^{-2}$ | 0.27 | Bland (2013) |
| | Transition simple-to-complex craters | $D_c$ | km | 10 | Cheng & Klimczak (2022) |
| Vesta | Right ascension of the spin axis | $\alpha_0$ | deg. | $309.03 \pm 0.01$ | Russell et al. (2012b) |
| | Declination of the spin axis | $\delta_0$ | deg. | $42.23 \pm 0.01$ | Russell et al. (2012b) |
| | Mean radius | $R$ | km | $262.7 \pm 0.1$ | Russell et al. (2012b) |
| | Mean density | $\rho$ | kg.m$^{-3}$ | $3456 \pm 35$ | Schenk et al. (2021) |
| | Surface gravity | $g$ | m.s$^{-2}$ | 0.22 | Cheng & Klimczak (2022) |
| | Transition simple-to-complex craters | $D_c$ | km | 38 | Vincent et al. (2014) |
| Impactor | C-type impactor density | $\rho_{i,C}$ | kg.m$^{-3}$ | $1260 \pm 70$ | Chesley et al. (2014) |
| | S-type impactor density | $\rho_{i,S}$ | kg.m$^{-3}$ | $1950 \pm 140$ | Kanamaru & Sasaki (2019) |
| | Impactor angle | $\theta$ | deg. | 45 | |
| | Impactor speed | $v$ | m.s$^{-1}$ | 4790 | O'Brien & Sykes (2011) |

## 6. Discussion

The deficit of binary craters corresponding to binary asteroids with close-in small satellites (the bulk of the currently known population) is due to the absence of morphological signatures for impact synchronicity in such impact cases (Miljković et al., 2013). The separation/size ratio of the binary craters we have identified for Ceres and for Vesta (Fig. 4) is similar to that of the binary craters on Mars (Vavilov et al., 2022), and is different from the typical binary asteroid population (Margot et al., 2015). While the impact craters found by Vavilov et al. (2022)

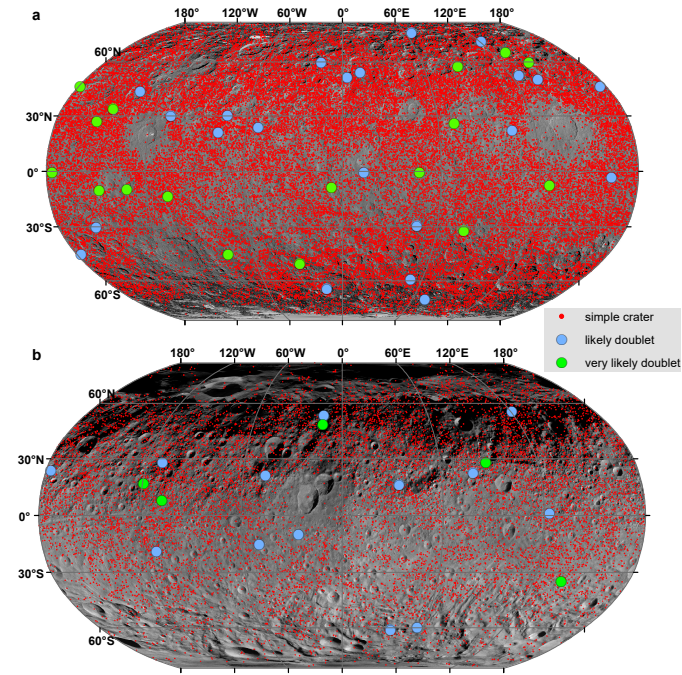

**Fig. 3.** Location of the identified binary craters on (a) Ceres and (b) Vesta. Simple craters from the crater databases (red dots) are distinguished from likely (blue circles) and very likely (green circles) binary craters.

on Mars are almost evenly distributed in the different morphological classes proposed by Miljković et al. (2013), classes of binary craters identified on both asteroids are limited, which is mostly due to the lithological characteristics of their surfaces, which prevent the formation of a thick and continuous ejecta blanket.

Compared to the current census of binary asteroids observed (MBAs, MCAs, and NEAs), which are mostly pairs of asteroids that have a small semi-major axis and are significantly different in size, our proposed synchronous impact craters observed on Ceres and Vesta (as also observed on Mars by Vavilov et al., 2022) suggest that they could have been formed by a population of binary asteroids that have not yet been observed. Moreover, the presence of morphological characteristics indicating synchronous impacts suggests the relative freshness of this crater population. Therefore, the binary asteroid population inferred here corresponds to contemporaneous systems that are comparable to systems known and observed today, and does not correspond to an ancient population, which is likely to have disappeared.

The distribution of the orientations of the binary craters in our catalogs is not anisotropic. We compare the orientation of the binary crater with our numerical simulations in Fig. 5. We performed a Kolmogorov-Smirnov test and found that for values of significance level $\alpha$ of between 0.01 and 0.2, the value of $D_{statistic}$ resulting from the test is always higher than this critical value $\alpha$, and consequently there is a significant difference between the simulations and the observations. Choosing a smaller value of $\alpha$ is not an option as it could lead to a greater number of type II errors. These results are also in agreement with the findings of Vavilov et al. (2022) for binary craters on planetary surfaces, which suggest they cannot be explained by a population of binary asteroids with zero obliquity.

We compared our results with a previous study that also explored the possible existence of binary craters on Ceres and Vesta

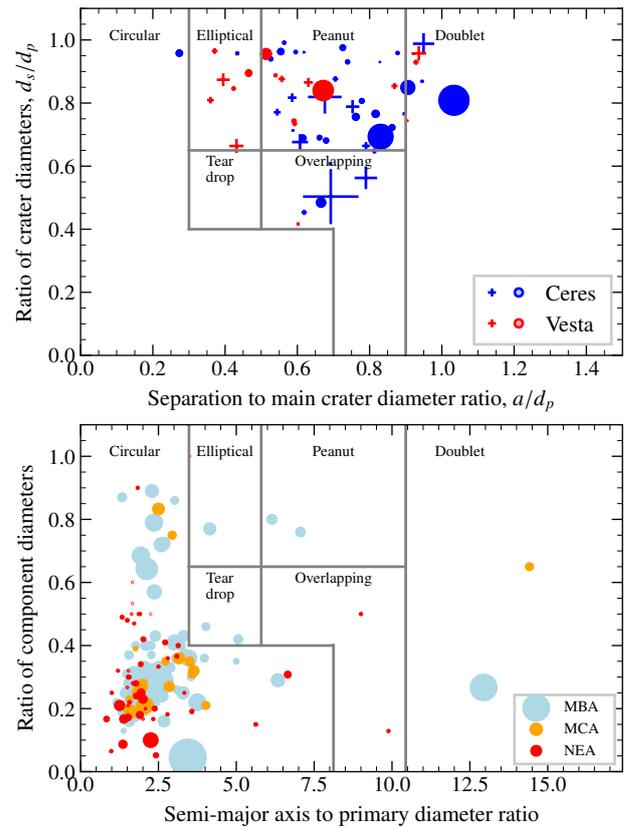

**Fig. 4.** Comparison between the population of binary craters (top) associated with the impact of binary asteroids, and the observed population of binary asteroids (bottom). Top: Distribution of the craters in terms of the separation–size ratio space for Ceres (blue) and Vesta (red). Crosses represent likely binary craters while the circles are the very likely binary craters (see Sect. 3). Symbol size is indicative of the primary crater diameter. We note that the absence of data in the left part of the graph is a consequence of the limitation of our survey and likely does not represent the absence of binary craters close to each other. Bottom: Distribution of binary asteroids in semimajor axis/size ratio space, color-coded according to their orbital characteristics: MBAs in light blue, MCAs in yellow, and NEAs in red. The morphological crater classification defined by Miljković et al. 2013 is added in both plots as a guideline.

Wren & Fevig (2018, 2023). Although this latter study was restricted to craters of 3 km in size or larger and to a smaller area —covering an area of 430,000 km² on Ceres (15%) and 25,000 km² on Vesta (3%) near their equators—, the authors identified some binary craters and found that 0.7% of impact events on Ceres and 2.1% of impact events on Vesta were binary impacts. In our global study, we find significantly fewer craters formed by binary asteroids: 0.31 % and 0.175% for Ceres and Vesta, respectively. This discrepancy might be explained by a difference of morphological interpretation: Wren & Fevig (2018) and Wren & Fevig (2023) seem to have included degraded and ancient craters in their binary crater catalogs for which morphological signatures are more subtle, and might lead to errors. As opposed to Wren & Fevig (2023), the aim of the present study is not to estimate the portion of binary impacts on planetary bodies but rather their size and orbital properties, and therefore these differences do not influence our conclusions. We thus strived for purity at the cost of the completeness of the present catalog.

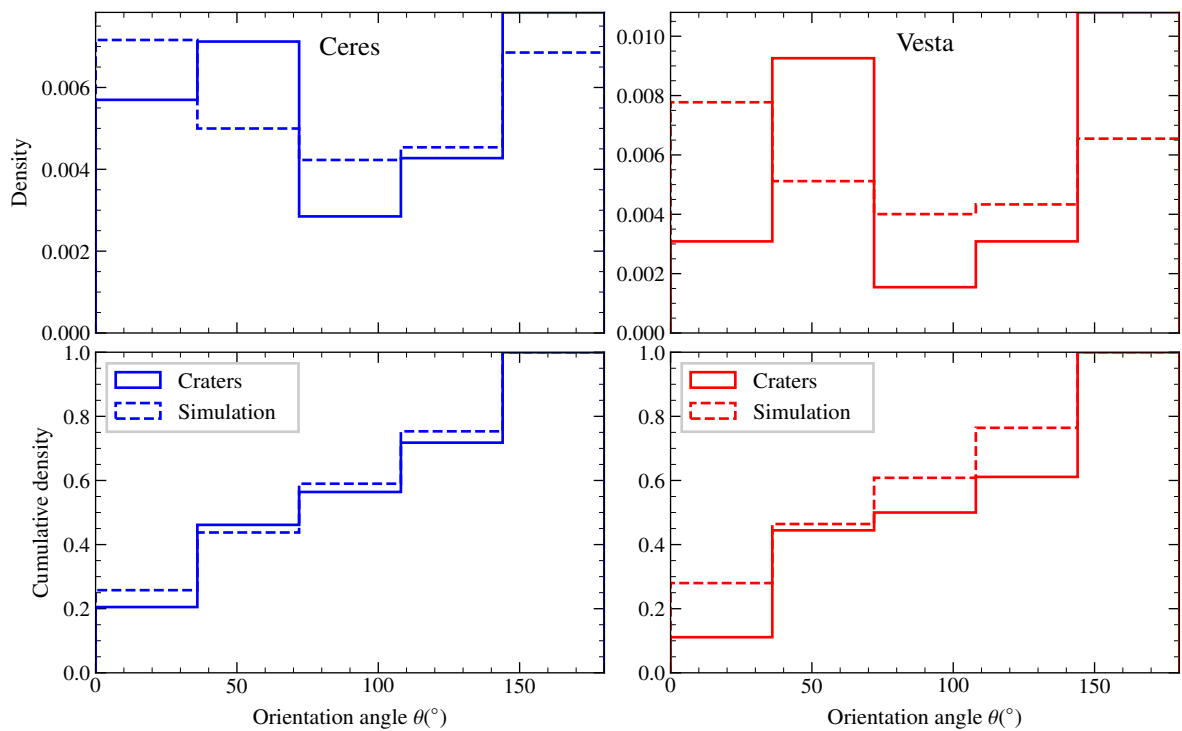

**Fig. 5.** Relative (top) and cumulative (bottom) distributions of the identified binary craters compared with the numerical simulations for Ceres (left, blue) and Vesta (right, red).

As already discussed by Vavilov et al. (2022), there are strong observational biases affecting the current census of binary asteroids. While radar detections are not affected by the orientation of the binary asteroid mutual plane, they are strongly limited in distance to the Earth, and hence in asteroid dynamical population. Conversely, light curves can effectively detect satellites around more distant populations, such as asteroids in the main belt; however, the probability of detection dramatically drops with increasing separation and obliquity of the systems (Vavilov et al., 2022).

## 7. Conclusions

We extended the recent work of Vavilov et al. (2022) on Martian binary craters to the asteroids Ceres and Vesta, targets of the NASA DAWN mission. We find evidence for synchronous crater formation related to the impact of 39 binary asteroids on Ceres, and 18 on Vesta. The components of the binary asteroids responsible for these impacts are nearly equal in size (diameter ratio above 0.5) and typically more separated (4–10 primary diameters), thus significantly differing from most known binary asteroids (Margot et al., 2015).

The distribution of the orientation of these binary craters on the surface appears isotropic, with no preferred direction, as found on Mars. This is striking as the obliquity of the three targets is widely different, from only 4° for Ceres to above 25° for Vesta and Mars. We compared this distribution of orientations with numerical simulations of the impact of binary asteroids whose mutual orbits are coplanar with their heliocentric orbits (reproducing the nonisotropic distribution of known binary systems; Pravec et al., 2006). While the sample of binary craters remains limited in size, we find a statistical mismatch between the predicted orientation and those measured.

These binary craters suggest the existence of a population of binary asteroids with properties in size ratio, mutual distance, and orientation that are different from those of the bulk of currently known binary systems. Considering the biases of the different techniques that allow the discovery of the satellites of asteroids (see Vavilov et al., 2022), and the recent discoveries of unexpected satellites (e.g., around Dinkinesh and Arecibo, during *Lucy* flyby and using *Gaia* astrometry; see Tanga et al., 2023), the current census of binary asteroid systems is likely biased. Additional observing efforts, using astrometry or stellar occultations for instance, may reveal satellites that have so far remained beyond the reach of direct imaging, light curves, and radar echoes (e.g., Pravec & Scheirich, 2012; Segev et al., 2023).

*Acknowledgements.* We thank an anonymous reviewer for the constructive comment on this manuscript. We thank the Lagrange laboratory of the Observatoire de la Côte d'Azur for supporting the work by C. Herrera. A.L. is funded by the Australian Research Council grant DP210100336, Curtin University, and the French government under the France 2030 investment plan, as part of the Initiative d'Excellence d'Aix-Marseille Université - A*MIDEX AMX-21-RID-O47. C. Herrera is currently a PhD student at the German Aerospace Center (DLR).

# Appendix A: Survey of binary craters on Ceres and Vesta

Table A.1. Catalog of binary craters on the surface of Ceres. For each, the ID, longitude, latitude and diameter of the two components are reported.

| | Crater, $c_1$ | | | | Crater, $c_2$ | | | | | Craters system | | | |
|---|---|---|---|---|---|---|---|---|---|---|---|---|---|
| ID | Lat (°) | Long (°) | Diam (km) | ID | Lat (°) | Long (°) | Diam (km) | Distance (km) | Diameter ratio | Separation ratio | Class | Azimuth (°) | Confidence |
| 0467-003 | -0.34 | 46.80 | 1.98 | 0468-004 | -0.42 | 46.90 | 1.86 | 1.04 | 0.94 | 0.53 | 5 | 37.36 | 2 |
| 0700+257 | 25.80 | 70.02 | 2.54 | 0699+260 | 26.03 | 69.93 | 2.05 | 1.97 | 0.81 | 0.78 | 5 | -109.83 | 2 |
| 0774-322 | -32.22 | 77.52 | 3.03 | 0778-323 | -32.33 | 77.81 | 2.95 | 2.20 | 0.98 | 0.73 | 5 | 24.63 | 2 |
| 1262-076 | -7.68 | 126.23 | 2.02 | 1264-076 | -7.66 | 126.42 | 1.88 | 1.49 | 0.93 | 0.74 | 5 | -5.56 | 2 |
| 1836-008 | -0.74 | -176.31 | 3.63 | 1838-005 | -0.65 | -176.24 | 3.48 | 0.99 | 0.96 | 0.27 | 1 | -50.22 | 2 |
| 1810+458 | 45.84 | -178.88 | 3.51 | 1813+456 | 45.61 | -178.70 | 2.42 | 2.16 | 0.69 | 0.61 | 5 | 59.78 | 2 |
| 2060+268 | 26.84 | -153.94 | 4.13 | 2055+266 | 26.69 | -154.37 | 3.16 | 3.37 | 0.77 | 0.82 | 5 | 158.86 | 2 |
| 2114-103 | -10.35 | -148.66 | 3.12 | 2110-103 | -10.30 | -148.99 | 2.25 | 2.69 | 0.72 | 0.86 | 5 | -171.13 | 2 |
| 2283-097 | -9.71 | -131.62 | 3.54 | 2286-096 | -9.71 | -131.37 | 3.41 | 1.96 | 0.96 | 0.55 | 5 | -0.94 | 2 |
| 2129+338 | 33.84 | -147.05 | 2.36 | 2131+339 | 33.92 | -146.84 | 1.63 | 1.56 | 0.69 | 0.66 | 5 | -24.98 | 2 |
| 2529-138 | -13.81 | -107.04 | 3.91 | 2526-136 | -13.59 | -107.34 | 2.95 | 2.98 | 0.76 | 0.76 | 5 | -143.85 | 2 |
| 2825-450 | -45.06 | -77.50 | 1.38 | 2824-449 | -44.92 | -77.56 | 1.33 | 1.21 | 0.96 | 0.88 | 5 | -105.84 | 2 |
| 3533-085 | -8.55 | -6.66 | 5.51 | 3532-089 | -8.97 | -6.82 | 2.67 | 3.67 | 0.48 | 0.67 | 4 | 110.51 | 2 |
| 3301-502 | -50.32 | -29.92 | 8.64 | 3312-497 | -49.70 | -28.79 | 7.34 | 7.83 | 0.85 | 0.91 | 6 | -39.85 | 2 |
| 3076+235 | 23.58 | -52.40 | 3.56 | 3079+237 | 23.80 | -52.10 | 2.31 | 2.90 | 0.65 | 0.81 | 4 | -38.67 | 2 |
| 2871+303 | 30.34 | -72.86 | 1.48 | 2870+302 | 30.29 | -72.97 | 1.42 | 0.92 | 0.96 | 0.62 | 5 | 151.60 | 1 |
| 2830+207 | 20.85 | -76.98 | 3.33 | 2829+211 | 21.09 | -77.02 | 3.20 | 1.98 | 0.96 | 0.60 | 5 | -99.11 | 1 |
| 2512+299 (c) | 29.96 | -108.70 | 1.69 | 2512+299 | 29.97 | -108.84 | 1.20 | 0.99 | 0.71 | 0.59 | 5 | -176.01 | 1 |
| 2240+433 (c) | 43.37 | -135.92 | 1.19 | 2240+433 | 43.33 | -135.76 | 1.10 | 0.98 | 0.93 | 0.83 | 5 | 16.29 | 1 |
| 2039-303 | -30.32 | -156.07 | 5.34 | 2033-301 | -30.14 | -156.64 | 3.55 | 4.22 | 0.66 | 0.79 | 5 | -160.49 | 1 |
| 1761+462 | 46.25 | 176.22 | 1.31 | 1761+462 (c) | 46.11 | 176.31 | 1.25 | 1.21 | 0.96 | 0.92 | 6 | 66.45 | 1 |
| 1829-448 | -44.90 | -177.02 | 4.75 | 1834-449 | -44.86 | -176.58 | 3.66 | 2.59 | 0.77 | 0.54 | 5 | -7.74 | 1 |
| 1636-031 | -3.10 | 163.61 | 2.12 | 1635-033 | -3.33 | 163.51 | 1.84 | 2.01 | 0.87 | 0.95 | 6 | 114.68 | 1 |
| 0129-009 | -0.88 | 12.94 | 2.30 | 0127-007 | -0.77 | 12.78 | 1.40 | 1.59 | 0.61 | 0.69 | 4 | -144.55 | 1 |
| 0470-296 | -29.65 | 47.05 | 3.87 | 0470-300 | -29.98 | 47.05 | 3.39 | 2.73 | 0.88 | 0.71 | 5 | 90.56 | 1 |
| 1054+219 | 22.01 | 105.38 | 2.96 | 1054+219 (c) | 22.00 | 105.55 | 2.83 | 1.28 | 0.96 | 0.43 | 2 | 4.97 | 1 |
| 1368+499 | 49.92 | 136.89 | 1.97 | 1371+499 | 49.98 | 137.21 | 1.51 | 1.77 | 0.77 | 0.90 | 5 | -15.48 | 1 |
| 0857+572 | 57.38 | 85.72 | 16.03 | 0866+589 | 58.95 | 86.52 | 11.11 | 13.31 | 0.69 | 0.83 | 5 | -75.34 | 2 |
| 1313+658 | 65.85 | 131.56 | 19.54 | 1313+658 (c) | 68.00 | 134.61 | 15.80 | 20.19 | 0.81 | 1.03 | 6 | -62.29 | 2 |
| 1412+595 | 59.55 | 141.32 | 2.72 | 1410+593 | 59.34 | 141.14 | 1.85 | 1.85 | 0.68 | 0.68 | 5 | 114.26 | 2 |
| 1251+520 | 51.99 | 125.29 | 22.83 | 1263+503 | 50.23 | 126.35 | 18.70 | 15.43 | 0.82 | 0.68 | 5 | 68.96 | 1 |
| 0035+511 | 51.10 | 3.43 | 14.48 | 0009+511 | 51.08 | 0.76 | 14.31 | 13.75 | 0.99 | 0.95 | 6 | -179.84 | 1 |
| 0128+537 | 53.74 | 12.80 | 3.66 | 0123+536 | 53.68 | 12.34 | 1.66 | 2.26 | 0.45 | 0.62 | 4 | 168.16 | 1 |
| 3438+596 | 59.65 | -16.07 | 9.08 | 3431+604 | 60.40 | -16.83 | 7.16 | 6.84 | 0.79 | 0.75 | 5 | -116.57 | 1 |
| 1214+728 | 72.84 | 121.67 | 15.36 | 1238+741 | 74.15 | 124.11 | 8.64 | 12.13 | 0.56 | 0.79 | 4 | -63.27 | 1 |
| 0653+784 | 78.54 | 65.81 | 5.93 | 0635+787 | 78.77 | 63.99 | 4.85 | 3.47 | 0.82 | 0.59 | 5 | -146.71 | 1 |
| 0713-719 | -71.80 | 71.43 | 10.90 | 0688-717 | -71.64 | 68.91 | 7.37 | 6.62 | 0.68 | 0.61 | 5 | -170.02 | 1 |
| 3489-649 | -64.92 | -12.36 | 37.77 | 3404-665 | -66.57 | -19.01 | 19.01 | 26.16 | 0.50 | 0.69 | 4 | 145.86 | 1 |
| 0513-593 | -59.31 | 51.48 | 3.24 | 0517-590 | -59.12 | 51.72 | 3.22 | 1.83 | 0.99 | 0.56 | 5 | -57.00 | 1 |

**Table A.2.** Catalog of binary craters on the surface of Vesta. For each, the ID, longitude, latitude and diameter of the two components are reported.

| | Crater, $c_1$ | | | Crater, $c_2$ | | | | Craters system | | | | |
|---|---|---|---|---|---|---|---|---|---|---|---|---|
| ID | Lat (°) | Long (°) | Diam (km) | ID | Lat (°) | Long (°) | Diam (km) | Distance (km) | Diameter ratio | Separation ratio | Class | Azimuth (°) | Confidence |
| 1210+554 | 55.45 | 121.03 | 6.12 | 1215+546 | 54.63 | 121.50 | 5.29 | 3.86 | 0.87 | 0.63 | 5 | 71.73 | 1 |
| 0128+531 | 53.16 | -12.87 | 9.63 | 0143+528 | 52.87 | -14.34 | 6.39 | 4.16 | 0.66 | 0.43 | 2 | 162.28 | 1 |
| 0129+482 | 48.21 | -12.96 | 12.99 | 0106+494 | 49.43 | -10.63 | 10.91 | 8.72 | 0.84 | 0.67 | 5 | -39.33 | 2 |
| 1073+079 | 7.92 | -107.32 | 1.48 | 1074+080 | 8.02 | -107.42 | 1.26 | 0.63 | 0.85 | 0.42 | 2 | -135.33 | 2 |
| 1194+169 | 16.99 | -119.49 | 3.55 | 1193+173 | 17.32 | -119.31 | 3.17 | 1.65 | 0.89 | 0.47 | 2 | -63.02 | 2 |
| 0463+209 | 20.96 | -46.33 | 3.68 | 0456+213 | 21.33 | -45.61 | 3.42 | 3.41 | 0.93 | 0.93 | 6 | -29.00 | 1 |
| 1771+236 | 23.63 | -177.18 | 3.97 | 1768+238 | 23.81 | -176.88 | 3.83 | 1.47 | 0.97 | 0.37 | 2 | -32.81 | 1 |
| 1103+278 | 27.84 | -110.31 | 4.12 | 1112+279 | 27.92 | -111.22 | 3.52 | 3.58 | 0.85 | 0.87 | 5 | -173.97 | 1 |
| 1118-188 | -18.89 | -111.86 | 4.57 | 1116-186 | -18.60 | -111.61 | 3.70 | 1.64 | 0.81 | 0.36 | 2 | -50.14 | 1 |
| 0498-152 | -15.29 | -49.84 | 8.82 | 0492-147 | -14.71 | -49.30 | 7.71 | 3.48 | 0.87 | 0.39 | 2 | -47.86 | 1 |
| 0260-098 | -9.89 | -26.05 | 10.01 | 0281-098 | -9.82 | -28.19 | 9.58 | 9.37 | 0.96 | 0.94 | 6 | -178.00 | 1 |
| 0882+279 | 27.92 | 88.22 | 2.08 | 0879+279 | 28.00 | 87.92 | 1.54 | 1.23 | 0.74 | 0.59 | 5 | -163.30 | 2 |
| 1230+010 | 1.10 | 123.07 | 2.93 | 1227+009 | 0.95 | 122.75 | 2.60 | 1.58 | 0.89 | 0.54 | 5 | 155.44 | 1 |
| 0795+223 | 22.32 | 79.58 | 4.46 | 0795+228 | 22.88 | 79.58 | 3.91 | 2.48 | 0.88 | 0.56 | 5 | -89.88 | 1 |
| 0557-599 | -59.93 | 55.73 | 2.19 | 0553-597 | -59.73 | 55.31 | 0.91 | 1.32 | 0.42 | 0.60 | 4 | -136.56 | 1 |
| 0364-615 | -61.54 | 36.43 | 3.34 | 0355-613 | -61.37 | 35.56 | 2.45 | 1.98 | 0.73 | 0.59 | 5 | -158.58 | 1 |
| 1377-350 | -35.02 | 137.80 | 6.52 | 1385-354 | -35.49 | 138.52 | 6.23 | 3.35 | 0.96 | 0.51 | 5 | 38.52 | 2 |
| 0342+160 | 16.00 | 34.23 | 1.34 | 0339+161 | 16.13 | 33.98 | 1.00 | 1.21 | 0.74 | 0.90 | 6 | -151.23 | 1 |

Table A.3. Impactor sizes of the binary craters of Ceres, estimated for two asteroid compositions, C-type and S-type.

| | Main crater, $c_1$ | | | | Secondary crater, $c_2$ | | | | Impactor diameters | | | |
|---|---|---|---|---|---|---|---|---|---|---|---|---|
| ID | Lat (°) | Long (°) | Diam (km) | ID | Lat (°) | Long (°) | Diam (km) | $c_1$ (S-type) (km) | $c_1$ (C-type) (km) | $c_2$ (S-type) (km) | $c_2$ (C-type) (km) |
| 0467-003 | -0.34 | 46.80 | 1.98 | 0468-004 | -0.42 | 46.90 | 1.86 | 0.10 | 0.18 | 0.10 | 0.16 |
| 0700+257 | 25.80 | 70.02 | 2.54 | 0699+260 | 26.03 | 69.93 | 2.05 | 0.14 | 0.23 | 0.11 | 0.18 |
| 0774-322 | -32.22 | 77.52 | 3.03 | 0778-323 | -32.33 | 77.81 | 2.95 | 0.17 | 0.28 | 0.16 | 0.28 |
| 1262-076 | -7.68 | 126.23 | 2.02 | 1264-076 | -7.66 | 126.42 | 1.88 | 0.11 | 0.18 | 0.10 | 0.17 |
| 1836-008 | -0.74 | -176.31 | 3.63 | 1838-005 | -0.65 | -176.24 | 3.48 | 0.21 | 0.35 | 0.20 | 0.33 |
| 1810+458 | 45.84 | -178.88 | 3.51 | 1813+456 | 45.61 | -178.70 | 2.42 | 0.20 | 0.34 | 0.13 | 0.22 |
| 2060+268 | 26.84 | -153.94 | 4.13 | 2055+266 | 26.69 | -154.37 | 3.16 | 0.24 | 0.41 | 0.18 | 0.30 |
| 2114-103 | -10.35 | -148.66 | 3.12 | 2110-103 | -10.30 | -148.99 | 2.25 | 0.17 | 0.29 | 0.12 | 0.20 |
| 2283-097 | -9.71 | -131.62 | 3.54 | 2286-096 | -9.71 | -131.37 | 3.41 | 0.20 | 0.34 | 0.19 | 0.33 |
| 2129+338 | 33.84 | -147.05 | 2.36 | 2131+339 | 33.92 | -146.84 | 1.63 | 0.13 | 0.21 | 0.08 | 0.14 |
| 2529-138 | -13.81 | -107.04 | 3.91 | 2526-136 | -13.59 | -107.34 | 2.95 | 0.22 | 0.38 | 0.16 | 0.28 |
| 2825-450 | -45.06 | -77.50 | 1.38 | 2824-449 | -44.92 | -77.56 | 1.33 | 0.07 | 0.12 | 0.07 | 0.11 |
| 3533-085 | -8.55 | -6.66 | 5.51 | 3532-089 | -8.97 | -6.82 | 2.67 | 0.33 | 0.56 | 0.15 | 0.25 |
| 3301-502 | -50.32 | -29.92 | 8.64 | 3312-497 | -49.70 | -28.79 | 7.34 | 0.55 | 0.94 | 0.46 | 0.78 |
| 3076+235 | 23.58 | -52.40 | 3.56 | 3079+237 | 23.80 | -52.10 | 2.31 | 0.20 | 0.34 | 0.12 | 0.21 |
| 2871+303 | 30.34 | -72.86 | 1.48 | 2870+302 | 30.29 | -72.97 | 1.42 | 0.07 | 0.13 | 0.07 | 0.12 |
| 2830+207 | 20.85 | -76.98 | 3.33 | 2829+211 | 21.09 | -77.02 | 3.20 | 0.19 | 0.32 | 0.18 | 0.30 |
| 2512+299 (c) | 29.96 | -108.70 | 1.69 | 2512+299 | 29.97 | -108.84 | 1.20 | 0.09 | 0.15 | 0.06 | 0.10 |
| 2240+433 (c) | 43.37 | -135.92 | 1.19 | 2240+433 | 43.33 | -135.76 | 1.10 | 0.06 | 0.10 | 0.05 | 0.09 |
| 2039-303 | -30.32 | -156.07 | 5.34 | 2033-301 | -30.14 | -156.64 | 3.55 | 0.32 | 0.54 | 0.20 | 0.34 |
| 1761+462 | 46.25 | 176.22 | 1.31 | 1761+462 (c) | 46.11 | 176.31 | 1.25 | 0.06 | 0.11 | 0.06 | 0.10 |
| 1829-448 | -44.90 | -177.02 | 4.75 | 1834-449 | -44.86 | -176.58 | 3.66 | 0.28 | 0.48 | 0.21 | 0.35 |
| 1636-031 | -3.10 | 163.61 | 2.12 | 1635-033 | -3.33 | 163.51 | 1.84 | 0.11 | 0.19 | 0.10 | 0.16 |
| 0129-009 | -0.88 | 12.94 | 2.30 | 0127-007 | -0.77 | 12.78 | 1.40 | 0.12 | 0.21 | 0.07 | 0.12 |
| 0470-296 | -29.65 | 47.05 | 3.87 | 0470-300 | -29.98 | 47.05 | 3.39 | 0.22 | 0.38 | 0.19 | 0.32 |
| 1054+219 | 22.01 | 105.38 | 2.96 | 1054+219 (c) | 22.00 | 105.55 | 2.83 | 0.16 | 0.28 | 0.16 | 0.26 |
| 1368+499 | 49.92 | 136.89 | 1.97 | 1371+499 | 49.98 | 137.21 | 1.51 | 0.10 | 0.18 | 0.08 | 0.13 |
| 0857+572 | 57.38 | 85.72 | 16.03 | 0866+589 | 58.95 | 86.52 | 11.11 | 1.07 | 1.89 | 0.74 | 1.25 |
| 1313+658 | 65.85 | 131.56 | 19.54 | 1313+658 (c) | 68.00 | 134.61 | 15.80 | 1.38 | 2.36 | 1.05 | 1.86 |
| 1412+595 | 59.55 | 141.32 | 2.72 | 1410+593 | 59.34 | 141.14 | 1.85 | 0.15 | 0.25 | 0.10 | 0.16 |
| 1251+520 | 51.99 | 125.29 | 22.83 | 1263+503 | 50.23 | 126.35 | 18.70 | 1.69 | 2.82 | 1.31 | 2.25 |
| 0035+511 | 51.10 | 3.43 | 14.48 | 0009+511 | 51.08 | 0.76 | 14.31 | 0.94 | 1.68 | 0.93 | 1.66 |
| 0128+537 | 53.74 | 12.80 | 3.66 | 0123+536 | 53.68 | 12.34 | 1.66 | 0.21 | 0.35 | 0.08 | 0.14 |
| 3438+596 | 59.65 | -16.07 | 9.08 | 3431+604 | 60.40 | -16.83 | 7.16 | 0.59 | 0.99 | 0.45 | 0.76 |
| 1214+728 | 72.84 | 121.67 | 15.36 | 1238+741 | 74.15 | 124.11 | 8.64 | 1.02 | 1.80 | 0.55 | 0.94 |
| 0653+784 | 78.54 | 65.81 | 5.93 | 0635+787 | 78.77 | 63.99 | 4.85 | 0.36 | 0.61 | 0.29 | 0.49 |
| 0713-719 | -71.80 | 71.43 | 10.90 | 0688-717 | -71.64 | 68.91 | 7.37 | 0.72 | 1.22 | 0.46 | 0.78 |
| 3489-649 | -64.92 | -12.36 | 37.77 | 3404-665 | -66.57 | -19.01 | 19.01 | 3.22 | 4.99 | 1.34 | 2.29 |
| 0513-593 | -59.31 | 51.48 | 3.24 | 0517-590 | -59.12 | 51.72 | 3.22 | 0.18 | 0.31 | 0.18 | 0.30 |

Table A.4. Impactor sizes of the binary craters of Vesta, estimated for two asteroid compositions, C-type and S-type.

| | Main crater, $c_1$ | | | Secondary crater, $c_2$ | | | | Impactor diameters | | | |
|---|---|---|---|---|---|---|---|---|---|---|---|
| ID | Lat (°) | Long (°) | Diam (km) | ID | Lat (°) | Long (°) | Diam (km) | $c_1$ (S-type) (km) | $c_1$ (C-type) (km) | $c_2$ (S-type) (km) | $c_2$ (C-type) (km) |
| 1210+554 | 55.45 | 121.03 | 6.12 | 1215+546 | 54.63 | 121.50 | 5.29 | 0.37 | 0.63 | 0.32 | 0.54 |
| 0128+531 | 53.16 | -12.87 | 9.63 | 0143+528 | 52.87 | -14.34 | 6.39 | 0.63 | 1.06 | 0.39 | 0.66 |
| 0129+482 | 48.21 | -12.96 | 12.99 | 0106+494 | 49.43 | -10.63 | 10.91 | 0.82 | 1.49 | 0.72 | 1.22 |
| 1073+079 | 7.92 | -107.32 | 1.48 | 1074+080 | 8.02 | -107.42 | 1.26 | 0.07 | 0.13 | 0.06 | 0.10 |
| 1194+169 | 16.99 | -119.49 | 3.55 | 1193+173 | 17.32 | -119.31 | 3.17 | 0.20 | 0.34 | 0.18 | 0.30 |
| 0463+209 | 20.96 | -46.33 | 3.68 | 0456+213 | 21.33 | -45.61 | 3.42 | 0.21 | 0.36 | 0.19 | 0.33 |
| 1771+236 | 23.63 | -177.18 | 3.97 | 1768+238 | 23.81 | -176.88 | 3.83 | 0.23 | 0.39 | 0.22 | 0.37 |
| 1103+278 | 27.84 | -110.31 | 4.12 | 1112+279 | 27.92 | -111.22 | 3.52 | 0.24 | 0.40 | 0.20 | 0.34 |
| 1118-188 | -18.89 | -111.86 | 4.57 | 1116-186 | -18.60 | -111.61 | 3.70 | 0.27 | 0.45 | 0.21 | 0.36 |
| 0498-152 | -15.29 | -49.84 | 8.82 | 0492-147 | -14.71 | -49.30 | 7.71 | 0.57 | 0.96 | 0.49 | 0.82 |
| 0260-098 | -9.89 | -26.05 | 10.01 | 0281-098 | -9.82 | -28.19 | 9.58 | 0.65 | 1.11 | 0.62 | 1.05 |
| 0882+279 | 27.92 | 88.22 | 2.08 | 0879+279 | 28.00 | 87.92 | 1.54 | 0.11 | 0.19 | 0.08 | 0.13 |
| 1230+010 | 1.10 | 123.07 | 2.93 | 1227+009 | 0.95 | 122.75 | 2.60 | 0.16 | 0.27 | 0.14 | 0.24 |
| 0795+223 | 22.32 | 79.58 | 4.46 | 0795+228 | 22.88 | 79.58 | 3.91 | 0.26 | 0.44 | 0.22 | 0.38 |
| 0557-599 | -59.93 | 55.73 | 2.19 | 0553-597 | -59.73 | 55.31 | 0.91 | 0.12 | 0.20 | 0.04 | 0.07 |
| 0364-615 | -61.54 | 36.43 | 3.34 | 0355-613 | -61.37 | 35.56 | 2.45 | 0.19 | 0.32 | 0.13 | 0.22 |
| 1377-350 | -35.02 | 137.80 | 6.52 | 1385-354 | -35.49 | 138.52 | 6.23 | 0.40 | 0.68 | 0.38 | 0.65 |
| 0342+160 | 16.00 | 34.23 | 1.34 | 0339+161 | 16.13 | 33.98 | 1.00 | 0.07 | 0.11 | 0.05 | 0.08 |